\documentclass[a4paper,11pt]{article}
\usepackage[latin1]{inputenc}
\usepackage{graphicx}
\usepackage{bm}
\usepackage{epsfig}
\usepackage{amsmath}
\usepackage{amsfonts}
\usepackage{amssymb, longtable}
\usepackage{color}
\usepackage[round,authoryear]{natbib}
\usepackage{multirow}
\usepackage{amsmath, amsthm, amssymb}

\newtheorem{theorem}{Theorem}
\newtheorem{definition}{Definition}

\begin{document}

\title{\textbf{A spatial scan statistic for zero-inflated Poisson process}}

\author{\textbf{Andr\'e L. F. Can\c{c}ado, Cibele Queiroz da-Silva and Michel Ferreira da Silva}\\
Departamento de Estat\'{\i}stica,\\
Universidade de Bras\'ilia - UnB,\\
70910-900, Bras\'ilia, DF, Brazil\\
e-mail:\emph{cancado@gmail.com; cibeleqs@gmail.com; ferreirasilva.michel@gmail.com}
}

\maketitle
\vspace*{0.5cm}

\maketitle

\begin{abstract}
The scan statistic is widely used in spatial cluster detection applications of inhomogeneous Poisson processes. However, real data may present substantial departure from the underlying Poisson process. One of the possible departures has to do with zero excess. Some studies point out that when applied to data with excess zeros, the spatial scan statistic may produce biased inferences. In this work, we develop a closed-form scan statistic for cluster detection of spatial zero-inflated count data. We apply our methodology to simulated and real data. Our simulations revealed that the Scan-Poisson statistic steadily deteriorates as the number of zeros increases, producing biased inferences. On the other hand, our proposed Scan-ZIP and Scan-ZIP+EM statistics are, most of the time, either superior or comparable to the Scan-Poisson statistic.
\end{abstract}

\section{Introduction}
The spatial scan statistic proposed by \cite{KULL:97} has been widely used in spatial disease or crime surveillance and other spatial cluster detection applications. One of the versions of the scan statistic was developed for inhomogeneous Poisson processes. However, in practice, data may present substantial departure from the assumed underlying Poisson process.

The Poisson distribution is often used for analysis of count data. However, this  distribution presents some limitations. One of the most well known  is the fact that the variance of the distribution is equal to the mean.

Count data, like contingency tables, often contain cells having zero counts. However, a zero count is called a {\it sampling zero} when a zero is drawn from a distribution with positive mass at zero. According to \cite{Agresti:90}, a zero for a cell in which it is theoretically impossible to have observations is called a {\it structural zero}. \cite{Agresti:90} also remarks ``a sampling zero is an  observation having value 0, and we regard it as one of the observed counts. A structural zero is not an observation, however.''

\cite{OF:07} worked with regression models for count data related to zoological data. In their application, the number of {\it C. caretta} hatchlings dying from exposure to the sun was also modeled. The data included both structural zeros and sampling zeros. If no {\it C. caretta} hatchlings emerge from the nests, the number of {\it C. caretta} hatchlings dying from exposure to the sun was automatically zero. If {\it C. caretta} hatchlings emerged from the nests, the number of {\it C. caretta} hatchlings dying from exposure to the sun could be zero or greater than zero. In the application, since there are two kinds of zero counts, the data is considered to be zero-inflated and the Poisson model is no longer good enough to correctly predict the occurrence of zero counts. 

\cite{Lambert:92} introduced the zero-inflated Poisson (ZIP) regression model to account for excess zeros in counts of manufacturing defects. The model has been applied to inumerous situations. \cite{BDSMK:99} used a zero-inflated Poisson model in dental epidemiology data. 

Several studies have dealt with spatial count data with excess zeros. \cite{AGCP:02} formulated a Bayesian zero-inflated regression model for spatial count data. The authors applied their methods to counts of nest burrows in a given area where more than 82$\%$ of the counts are 0. \cite{RF:06} remark that, in ecological surveys, count data often include excess zeros. It can occur either due to the inclusion of a habitat unsuitable to the species, or due to the limited ability of the species to disperse into all parts of the study region. The authors used a spatial zero-inflated Poisson regression model to determine an oak species range. In the context of the study of rare diseases, \cite{GRLQ:10} argue that, among other methods for the detection of disease clusters, the scan statistic by \cite{KULL:97} may not be suitable for problems involving very rare diseases. According to the authors, for diseases with very low prevalence, the number of cases may be very low and excess zeros may cause bias in the inferences. However, the authors argue that ``a likelihood ratio test similar to that of the spatial scan statistic may be difficult to develop in closed form.''

In this work, we build upon the developments in \cite{KULL:97} and propose two new scan statistics, denoted as Scan-\mbox{ZIP} and Scan-\mbox{ZIP}+EM, allowing for the occurrence of zero-inflated spatial data. Based on simulated data, we found that the Scan-Poisson statistic \citep{KULL:97} steadily deteriorates as the number of structural zeros increases, producing biased inferences. Moreover, when the data include structural zeros, the Scan-Poisson statistic is more likely to detect spurious significant clusters than the Scan-\mbox{ZIP} and Scan-\mbox{ZIP}+EM. 

This article is organized as follows. In Section 2, we review the Kulldorff scan statistic. The Scan-ZIP statistic is described in Section 3. Some theoretical properties are discussed in Section 4, and applications to simulated and real data are presented in Section 5. 

\section{Kulldorff scan statistic}
Consider an inhomogeneous Poisson point process over $k$ regions or locations in a study area. Let $x_i$ be the number of cases in region $i$ with corresponding at-risk population $n_i$ under unit-specific relative risk $\theta_i$ such that $x_i \sim \mbox{Poisson}(n_i\theta_i)$. Further, let $Z$ be a subset of the indexes $\{1,2,\ldots,k\}$, describing a given zone, which represent a putative cluster. Define $\mathcal{Z}$ as a collection of zones in the study area. 

\cite{KULL:97} formulated a scan statistic that compares the total number of case-counts in zone $Z$, $\displaystyle x_Z=\sum_{i \in Z}^{} x_i$, with the total number of counts outside $Z$, $\displaystyle x_{\overline{Z}}=\sum_{i \in \overline{Z}}^{} x_i$, given the corresponding population counts, that is, $n_Z=\displaystyle \sum_{i \in Z}^{} n_i$ and $n_{\overline{Z}}=\displaystyle \sum_{i \in \overline{Z}}^{} n_i$. Let $n=n_{\overline{Z}} + n_{Z}$ and $x=x_{\overline{Z}} + x_{Z}$, and assume that $\theta_i=\theta_Z$ for every region $i$ $\in$ $Z$ and that $\theta_i=\theta_0$ for every region $i$ $\notin$ $Z$. The hypotheses of interest are given by

\begin{eqnarray}
\label{hyp1}
\nonumber H_0: \theta_Z=\theta_0 \ \ \mbox{v.s.} \ \ H_a: \theta_Z>\theta_0,
\ \ Z \ \ \in \ \ \mathcal{Z},
\end{eqnarray}
where $H_0$ implies that there is a constant risk, while $H_a$ implies that there is at least one cluster defined by a zone $Z\  \in \ \mathcal{Z}$ such that $\theta_Z > \theta_0$.  

Since the putative cluster $Z$ is actually unknown, it is considered a parameter, and the likelihood function $L(Z)=L(Z,\theta_0,\theta_Z)$ is given by
\begin{eqnarray}
\label{lik0}
\hspace*{-0.5cm}L(Z)\!\!&=&\!\!L(Z,\theta_0,\theta_Z)\!\!=\!\!
\left[ \displaystyle \prod_{i \in Z}^{} \frac{e^{-n_i\theta_Z}(n_i\theta_Z)^{x_i}}{x_i!} \right]\!\!
\left[ \displaystyle \prod_{j \notin Z}^{} \frac{e^{-n_j\theta_0}(n_j\theta_0)^{x_j}}{x_j!}\right].
\end{eqnarray}
The maximum likelihood estimators of $\theta_0$ and $\theta_Z$ are, respectively, $\hat{\theta}_0=x_{\overline{Z}}/n_{\overline{Z}}$ and $\hat{\theta}_Z=x_{Z}/n_{Z}$, and the most likely cluster is the solution of $\hat{Z}=\{Z: L(Z) \geq L(Z^{'}) \ \ \forall \ \ Z^{'} \in \mathcal{Z}\}$ where $\hat{Z}$ is the maximum likelihood estimator of parameter $Z$.

Besides showing how to find the most likely cluster, \cite{KULL:97} also developed a likelihood ratio test that allows us to decide whether or not the most likely cluster is statistically significant, that is, if the area included in the most likely cluster really incorporates an abnormally high number of cases.

For a given zone $Z$, we have

\begin{eqnarray}
\nonumber \lambda_Z &=&\frac{\mbox{sup}_{\ \theta_Z > \theta_0}L(Z,\theta_0,\theta_Z)}
{\mbox{sup}_{\ \theta_Z=\ \theta_0}L(Z,\theta_0,\theta_Z)}\\
\nonumber & &  \\
&=& \left(\frac{x_Z/n_Z}{x/n}\right)^{x_Z}\left(\frac{x_{\overline{Z}}/n_{\overline{Z}}}{x/n}\right)^{x_{\overline{Z}}}
\times I\left( x_Z/n_Z > x_{\overline{Z}}/n_{\overline{Z}}\right).
\end{eqnarray}

Kulldorff's scan statistic is described by
\begin{eqnarray}
\label{Kulld}
 \lambda &=& \sup_{Z}\lambda_Z=\frac{\mbox{sup}_{\ Z \in \mathcal{Z}, \ \theta_Z > \theta_0}L(Z,\theta_0,\theta_Z)}
{\mbox{sup}_{\ Z \in \mathcal{Z}, \ \theta_Z= \theta_0}L(Z,\theta_0,\theta_Z)}.
\end{eqnarray}

\section{The Scan-\mbox{ZIP} statistic}

Now, in order to describe our scan statistic, consider the same notation as in Section 2, but now assume that the case-counts in the regions follow independent \mbox{ZIP} random variables with the same probability $p$ of a structural zero, that is, $X_i \sim \mbox{ZIP}(p, n_i\theta_i)$ where $X_i$s are independent. Thus, $P(X_i = x_i \mid p, n_i\theta_i) = P(X_i)$ is given by 

\begin{eqnarray}
\label{eq1}
\nonumber P(X_i=0 \mid p, n_i\theta_i) &=& p + (1-p)e^{-n_i\theta_i};\\
P(X_i=x_i \mid p, n_i\theta_i) &=& (1-p)\frac{e^{-n_i\theta_i}(n_i\theta_i)^{x_i}}{x_i!}, \ \ x_i>0.
\end{eqnarray}
The \mbox{ZIP} model allows for additional flexibility when compared to the Poisson model. If $X_i \sim \mbox{ZIP}(p, n_i\theta_i)$, then $E(X_i\mid p, n_i\theta_i)=(1-p)n_i\theta_i$ and $V(X_i\mid p, n_i\theta_i)=(1-p)n_i\theta_i(1+pn_i\theta_i)$.

It can be verified that $E(X_i \mid p, n_i\theta_i)<n_i\theta_i$ when $p>0$. Therefore, when structural zeros occur, the \mbox{ZIP} model correctly accounts, on average, for a reduction in the case-counts. Further, since $V(X_i\mid p, n_i\theta_i)>E(X_i \mid p, n_i\theta_i)$ when $p>0$, the \mbox{ZIP} model allows for overdispersion or extra-Poisson variation. The Poisson model often underestimates the observed dispersion.

Regarding the likelihood ratio test formulation for the \mbox{ZIP} model, assume, as usual for the Kulldorff scan statistic, that $\theta_i=\theta_Z$ for every region $i$ $\in$ ${Z}$ and that $\theta_j=\theta_0$ for every region $j$ $\notin$ ${Z}$. The hypotheses of interest are given by

\begin{eqnarray}
\label{hyp2}
\nonumber H_0: \theta_Z=\theta_0 \ \ \mbox{v.s.} \ \ H_a: \theta_Z>\theta_0,
\ \ Z \ \ \in \ \ \mathcal{Z}.
\end{eqnarray}
In the consequent sections, we describe our Scan-\mbox{ZIP} statistic in two settings: (a) when we know when a zero count is a structural one, and (b) when we do not know, for sure, whether or not a zero count is a structural one.

\subsection{Incomplete-data likelihood}
Assume that data include both, sampling and structural zeros, and that we are not able to tell the nature of each kind of zero. Then, the likelihood conditioned on zone $Z$ and the observed data $X$ is given by

\begin{eqnarray}
\label{eq2}
\hspace*{-0.3cm}\nonumber L(p, \theta_0, \theta_Z) = L(p, \theta_0,\theta_Z \mid Z, X)\!\!&=&\!\!
\left[ \displaystyle \prod_{i \in {Z}}^{} P(X_i) \right]\!\!
\left[ \displaystyle \prod_{i \notin {Z}}^{} P(X_i) \right],
\end{eqnarray}

where $X = (X_1, \ldots, X_k)$. Thus, the log-likelihood is given by

\scriptsize
\begin{eqnarray}
\nonumber \ell(p, \theta_0,\theta_Z)&=&\sum_{\scriptscriptstyle{\begin{array}{c} i \in {Z}\\ x_i=0\end{array}}} \log(P(X_i)) + 
\displaystyle \sum_{\begin{array}{c} i \in {Z}\\ x_i>0\end{array}} \log(P(X_i))\\
\nonumber && + \displaystyle \sum_{\begin{array}{c} j \notin {Z}\\x_j=0\end{array}}\log(P(X_j))+ 
\displaystyle \sum_{\begin{array}{c} j \notin {Z}\\x_j>0\end{array}} \log(P(X_j)) \\
\nonumber  & & \\
\nonumber &=& \displaystyle \sum_{\begin{array}{c} i \in {Z}\\ x_i=0\end{array}} \log[p+(1-p)e^{-n_i\theta_Z}]
+\displaystyle \sum_{\begin{array}{c} i \in {Z}\\ x_i>0\end{array}} [\log(1-p)-n_i\theta_Z + x_i\log(n_i\theta_Z)]\\
&& + \displaystyle \sum_{\begin{array}{c} j \notin {Z}\\ x_j=0\end{array}} \log[p+(1-p)e^{-n_j\theta_0}]
+ \displaystyle \sum_{\begin{array}{c} j \notin {Z}\\ x_j>0\end{array}} [\log(1-p)-n_j\theta_0 + x_j\log(n_j\theta_0)]. 
\end{eqnarray}
\normalsize

Using this likelihood, it is not possible to find closed form MLE estimators for the parameters. However, following the approach developed by \cite{Lambert:92}, we describe a new likelihood under the assumption that we know in advance the nature of each zero, whether structural or sampling.

\subsection{Complete-data likelihood}
Let $\delta=(\delta_1,\ldots,\delta_k)$ where $\delta_i$ is a binary variable that assumes $\delta_i=1$ for a structural zero count in region $i$. Thus, $\delta_i \sim \mbox{Bernoulli}(p)$. Consider that $\delta=(\delta_1,\ldots,\delta_k)$ is observed, so that now we work with bivariate data, that is, $(X_i,\delta_i), \ \ i=1,\ldots,k$. The likelihood function for zone $Z \in \mathcal{Z}$ is given by

\small
\begin{eqnarray}
\label{eq4}
&& \nonumber L(p, \theta_0,\theta_Z) =L(p, \theta_0,\theta_Z\mid Z, X, \delta) =
\!\!\! \left[ \displaystyle \prod_{i \in {Z}}^{} P(X_i,\delta_i) \right] \!\!\! \left[ \displaystyle \prod_{i \notin {Z}}^{} P(X_i,\delta_i) \right] \\
\nonumber &&\\     
\nonumber &=&\!\!\! \left[ \displaystyle \prod_{i \in {Z}}^{} P(X_i=x_i \mid \delta_i=d_i)P(\delta_i=d_i)\right]
\!\!\! \left[ \displaystyle \prod_{i \notin {Z}}^{} P(X_i=x_i \mid \delta_i=d_i)P(\delta_i=d_i)\right]\\
\nonumber &&\\ 
&=&\!\!\! \left[ \displaystyle \prod_{i \in {Z}}^{} p^{d_i} \!\! \left[(1-p)\frac{e^{-n_i\theta_Z}(n_i\theta_Z)^{x_i}}{x_i!}\right]^{(1-d_i)}\right]
\!\!\! \left[ \displaystyle \prod_{i \notin {Z}}^{} p^{d_i} \!\! \left[(1-p)\frac{e^{-n_i\theta_0}(n_i\theta_0)^{x_i}}{x_i!}\right]^{(1-d_i)}\right]\!,\!
\end{eqnarray}
\normalsize
since $P(X_i=0,\delta_i=1)=p$ \ and \ $P(X_i=x_i,\delta_i=0)=(1-p)\frac{e^{-n_i\theta_i}(n_i\theta_i)^{x_i}}{x_i!}$, $x_i\geq 0$. 
Notice that $P(X_i>0,\delta_i=1)=0$.

Under $H_a$, the likelihood is given by
\begin{eqnarray}
\label{eq5}
\nonumber L_{a}(p,\theta_0,\theta_Z)\!&=&\!
p^{\sum_{i=1}^{k}d_i}\ (1-p)^{k-\sum_{i=1}^{k}d_i} 
e^{-\theta_Z\sum_{i \in {Z}}^{}n_i(1-d_i)}\ 
\theta_Z^{\sum_{i \in {Z}}^{}x_i(1-d_i)}\\
\nonumber && \times e^{-\theta_0\sum_{i \notin {Z}}^{}n_i(1-d_i)}\ 
\theta_0^{\sum_{i \notin {Z}}^{}x_i(1-d_i)},
\end{eqnarray}
and, under $H_0$, it is as follows:
\begin{eqnarray}
\label{eq6}
\nonumber \hspace*{-1.0cm}L_{0}(p, \theta_0)\!&=&\! p^{\sum_{i=1}^{k}d_i}\ (1-p)^{k-\sum_{i=1}^{k}d_i}
\times e^{-\theta_0\sum_{i=1}^{k}n_i(1-d_i)}\ \theta_0^{\sum_{i=1}^{k}x_i(1-d_i)}.
\end{eqnarray}
Under $H_0$, the maximum likelihood estimators (MLEs) are
\begin{eqnarray}
\label{eq7}
\hat{\theta}_0=\frac{\sum_{i=1}^{k}x_i(1-d_i)}{\sum_{i=1}^{k}n_i(1-d_i)}\ \ \ \mbox{and}\ \ \
\hat{p}=\frac{\sum_{i=1}^{k}d_i}{k},
\end{eqnarray}
while, under $H_a$, the MLEs are
\begin{eqnarray}
\label{eq8}
\hat{\theta}_z=\frac{\sum_{i \in {Z}}^{}x_i(1-d_i)}
{\sum_{i \in {Z}}^{}n_i(1-d_i)}, 
\ \ \
\hat{\theta}_0=\frac{\sum_{j \notin {Z}}^{}x_j(1-d_j)}
{\sum_{j \notin {Z}}^{}n_j(1-d_j)}, 
\ \ \ \mbox{and}\ \ \ \hat{p}=\frac{\sum_{i=1}^{k}d_i}{k}.
\end{eqnarray}
Notice that, by the Factorization Theorem (see \cite{CasellaBerger:90}, pp. 250),   $\sum_{i=1}^{k}x_i(1-d_i)$, $\sum_{i=1}^{k}n_i(1-d_i)$, and $\sum_{i=1}^{k}d_i$ are jointly sufficient for $({\theta}_0,p)$, while $\sum_{i \in {Z}}^{}x_i(1-d_i)$, $\sum_{i \in {Z}}^{}n_i(1-d_i)$, $\sum_{j \notin {Z}}^{}x_j(1-d_j)$, $\sum_{j \notin {Z}}^{}n_j(1-d_j)$, and $\sum_{i=1}^{k}d_i$ are jointly sufficient for $({\theta}_z,{\theta}_0,p)$.

\subsection{Likelihood ratio test when $\delta$ is known}
\label{sec:known}
Let $L_{a}^{S}(p, \theta_0, \theta_Z\mid Z, X, \delta) = L_{a}(\hat{p}, \hat{\theta}_0,\hat{\theta}_Z\mid Z, X, \delta)$ and $L_{0}^{S}(p, \theta_0\mid X, \delta) =L_{0}(\hat{p}, \hat{\theta}_0 \mid X, \delta)$ be the likelihoods, under $H_a$ and under $H_0$, respectively, evaluated at their respective MLEs. Then, 

\begin{eqnarray}
\label{eq9}
\nonumber \lambda&=&\frac{\mbox{sup}_{Z \in \mathcal{Z}, \ \theta_Z > \theta_0} L(p,\theta_0,\theta_Z\mid Z, X,\delta)}
{\mbox{sup}_{Z \in \mathcal{Z}, \ \theta_Z= \theta_0} L(p,\theta_0,\theta_Z\mid Z, X,\delta)} =
\frac{\mbox{sup}_{Z \in \mathcal{Z}} L_{a}^{S}(p, \theta_0,\theta_Z \mid Z, X,\delta)}
{L_{0}^{S}(p,\theta_0 \mid X,\delta)}
\\ \nonumber 
\\ \nonumber 
\hspace*{-1.0cm} &=&\mbox{sup}_{Z \in \mathcal{Z}}
\frac{\left[\frac{\sum_{i \in {Z}}^{}x_i(1-d_i)}{\sum_{i \in {Z}}^{}n_i(1-d_i)}
\right]^{\sum_{i \in {Z}}^{}x_i(1-d_i)}
\left[\frac{\sum_{j \notin {Z}}^{}x_j(1-d_j)}{\sum_{j \notin {Z}}^{}n_j(1-d_j)}
\right]^{\sum_{j \notin {Z}}^{}x_j(1-d_j)}}
{\left[\frac{\sum_{i=1}^{k}x_i(1-d_i)}{\sum_{i=1}^{k}n_i(1-d_i)}
\right]^{\sum_{i=1}^{k}x_i(1-d_i)}}
\\
&& \times I\left(\frac{\sum_{i \in {Z}}^{}x_i(1-d_i)}{\sum_{i \in {Z}}^{}n_i(1-d_i)}
>\frac{\sum_{j \notin {Z}}^{}x_j(1-d_j)}{\sum_{j \notin {Z}}^{}n_j(1-d_j)}\right),
\end{eqnarray}
if there is at least one zone $Z$ such that $\frac{\sum_{i \in {Z}}^{}x_i(1-d_i)}{\sum_{i \in {Z}}^{}n_i(1-d_i)} >\frac{\sum_{j \notin {Z}}^{}x_j(1-d_j)}{\sum_{j \notin {Z}}^{}n_j(1-d_j)}$, and $\lambda=1$ otherwise.

As in the likelihood ratio test developed by Kulldorff (1997), it is not possible to find the exact distribution associated with $\lambda$. However, in the present case, in which the locations of excess zeros are known, we can just remove the sites with structural zeros from the data set and use the likelihood ratio test developed by Kulldorff. 

Using the EM algorithm, in the following sections, we describe a closed-form Scan-\mbox{ZIP} statistic for the case in which $\delta_i$s are not known.

\subsection{Likelihood ratio test when $\delta$ is unknown}
When $\delta$ is unknown, the likelihood ratio test for the Scan-\mbox{ZIP} statistic via the EM algorithm is given by
\\
E-step: We estimate $\delta_i$ by the conditional expectation of $\delta_i$ given $X_i$.
Since $(\delta_i \mid X_i) \sim \mbox{Bernoulli}(\zeta_i)$, then 
\begin{eqnarray}
\label{eq10}
\nonumber \zeta_i&=&E(\delta_i \mid X_i)=P(\delta_i=1 \mid X_i)\\  
\nonumber &=& \frac{P(X_i=x_i\mid \delta_i=1)P(\delta_i=1)}
{P(X_i=x_i\mid \delta_i=1)P(\delta_i=1)+P(X_i=x_i\mid \delta_i=0)P(\delta_i=0)}I(x_i=0) \\
\nonumber &=& \frac{p}{p+(1-p)e^{-n_i\theta_i}} I(x_i=0)\quad\\
&=& \left\{ \begin{array}{ll}
    \frac{p}{p+(1-p)e^{-n_i\theta_i}}, & x_i=0 \\[2mm]
    0,& x_i=1,2,\ldots
    \end{array}
     \right. .
\end{eqnarray}
Therefore, at the $m$-{\it th} iteration of the EM algorithm,
\begin{eqnarray}
\label{eq11}
\hat{\delta_i}^{(m)}=\frac{\hat{p}^{(m)}}{\hat{p}^{(m)}+(1-\hat{p}^{(m)})
e^{-n_i\hat{\theta}_i^{(m)}}}I(x_i=0), \ \ i=1,\ldots,k. 
\end{eqnarray}
M-step: Conditional on vector $\hat{\delta}^{(m)}=(\hat{\delta}_1^{(m)},\ldots,\hat{\delta}_k^{(m)})$, at the ($m$+1)-{\it th} iteration, the MLEs of $\theta_0$, $\theta_Z$, and $p$ are obtained according to the expressions given in (\ref{eq7}) (under $H_0$) and (\ref{eq8}) (under $H_a$), but with $d_i$ replaced by $\hat{\delta}_i^{(m)}$. Thus, considering $H_a$ we have
\begin{eqnarray}
\label{eq12}
\hat{\delta}_i^{(m+1)}&=& \left\{ \begin{array}{ll}
\frac{\hat{p}^{(m+1)}}{\hat{p}^{(m+1)}+(1-\hat{p}^{(m+1)})e^{-n_i\hat{\theta}_Z^{\ (m+1)}}},
& x_i=0 \ \ \mbox{and} \ \ i \in {Z}\\[3mm]
\frac{\hat{p}^{(m+1)}}{\hat{p}^{(m+1)}+(1-\hat{p}^{(m+1)})e^{-n_i\hat{\theta}_0^{\ (m+1)}}},
& x_i=0 \ \ \mbox{and} \ \ i \notin {Z}\\[3mm]
0,& x_i=1,2,\ldots
    \end{array}
     \right. .
\end{eqnarray}
The E-step and M-step are performed until convergence.

The likelihood ratio test for the case of unknown $\delta$ is also given by (\ref{eq9}), but with $d_i$ replaced by $\hat{\delta}_i^{(s)}$, where $s$ is the iteration in which convergence is achieved.  

Again, the distribution of $\lambda$ cannot be exactly determined and a Monte Carlo simulation is conducted to obtain critical value $\lambda^*$. As we do not know the exact number of structural zeros, nor their locations, the Monte Carlo replicas are constructed using  a parametric bootstrap-like framework. Thus, given $\hat{p}$, the estimate of $p$ for the observed data, each region of the map is randomly assigned as a structural zero location with probability $\hat{p}$. This  procedure is repeated $B$ times and will generate $B$ maps with as many structural zeros, on average, as it was estimated for the observed data. The estimated expected number of cases in each region, under $H_0$, will be  $\hat{E}_0(X_i)=(1-\hat{p})n_i\hat{\theta}_0$, where $\hat{p}$ and $\hat{\theta}_0$ are obtained as defined in (\ref{eq7}). Now, given the structural zeros in each of the $B$ bootstrap samples, we perform a Monte Carlo simulation. The procedure can be summarized in the following steps:

\begin{enumerate}
	\item For the original, data calculate the observed value of $\lambda$, that is, $\lambda_{obs}$, using any spatial cluster detection method;
	\item Using $\hat{p}$, randomly assign the structural zero regions;
	\item Obtain a replication of the original data set under $H_0$ by generating a multinomial random variable with total count given by $\sum_{i=1}^{k} n_i$ and probabilities proportional to $n_i\theta_0$ if the $i$-\textit{th} region was not assigned as a structural zero, and 0 otherwise, $i=1,\dots,k$;
	\item Calculate $\lambda$ for the generated data, that is, find the most likely cluster through the same detection method used in step 1 and compute its corresponding likelihood ratio value;
	\item Steps 2--4 are repeated $B$ times, generating $\lambda_1, \ldots, \lambda_B$;
	\item For a level of significance $\alpha$, calculate the $100(1-\alpha)\%$
percentile of $\lambda$s and call it $\lambda^{*}$. 
The rejection region is described by $R=\{w: w > \lambda^{*} \}$. We reject $H_0$
if $\lambda_{obs}$ $\in$ $R$.
\end{enumerate}

\section{Properties of the Scan-\mbox{ZIP} test statistic}
\subsection{Detection versus inference}
\cite{KULL:97} remarks that his formulation of the spatial scan statistic enables not only the location of the most likely cluster but also inferences related to it, in the sense that when the null hypothesis is rejected it is possible to locate the region that causes the rejection. According to the author, most of the statistical methods for cluster analysis of spatial point processes do not possess both of these features.

In his Theorem 1, \cite{KULL:97} shows that when the null hypothesis is rejected for a given data set, say $A$, whose most likely cluster is $\hat{Z}_A$, then any other data set that shares all the cases in $\hat{Z}_A$ also implies that $H_0$ will also be rejected regardless of the locations of the remaining cases in the alternative data set $\bar{A}$, provided that the total number of points in the map is preserved. Next, we present a new version of this theorem that has been generalized to the case of the Scan-\mbox{ZIP} statistic.

\begin{theorem} Consider an inhomogeneous Poisson point process observed over $k$ regions or locations in a study area. Let $D$ be a data set composed by triplets $\{(x_i,n_i,d_i)\}$ for which $\hat{Z}$, a subset of $\Omega=\{1, 2, \dots, k\}$, is the most likely cluster. Similarly, let $D'$ be an alternative data set such that $\sum_{i=1}^k x_i'(1-d_i')=\sum_{i=1}^k x_i(1-d_i)$, $\sum_{i=1}^{k}n_i(1-d_i)=\sum_{j=1}^{k}n_j'(1-d_j')$, and $n_i'=n_i$, for all $i$. If the null hypothesis is rejected under $D$ then it is also rejected under $D'$ whenever $\sum_{i \in \hat{Z}} x_i'(1-d_i')\geq \sum_{i \in \hat{Z}} x_i(1-d_i)$.
\end{theorem}

In other words, Theorem 1 states that, if population $n_i$ is kept unchanged for all regions in the map and the adjusted number of cases $\sum_{i\in \hat{Z}} x_i(1-d_i)$ inside $\hat{Z}$ is not decreased, then $\lambda(\hat{Z})$ cannot decrease. The proof for Theorem 1 can be found in the Appendix and consists of showing that the test statistic is an increasing function on $\sum_{i\in \hat{Z}} x_i(1-d_i)$, the adjusted number of cases inside the zone $\hat{Z}$.

\subsection{Power}
For the spatial scan statistic, \cite{KULL:97} defined the so-called \textit{individually most powerful} test (IMP) as a proxy for dealing with the limitations in the technique that prevents finding an \textit{uniformly most powerful} (UMP) test. As noted by Kulldorff, we cannot expect to find a UMP test except for the special case when there is only one cluster under the alternative hypothesis. The IMP tests are constructed as follows:
\\
(a) Partition the subset of the parameter space related to the composite alternative hypothesis, that is, $W=\{(Z, \theta_Z, \theta_0):\theta_Z>\theta_0, \ Z \ \in \ \mathcal{Z} \},$ into a countable number of sets $\{A_j\}$ to form a partition of $W$, that is, $A_i \cap A_j = \emptyset$ for all $i \neq j$ and $W=\cup_j A_{j}^{}$. Let $\Psi=\overline{W}\cup W$ represent the whole parameter space. 
\\
(b) Define a partition $\{R_j\}$ of the critical region $R$ (see Section 3.3 for the Scan-\mbox{ZIP} statistic).
\\
(c) Take an alternative critical region $R^{'}=\cup_j R^{'}_{j}$ in which $\{R^{'}_j\}$ represents a partition of $R^{'}$.

Let $\pi(\theta) = P(w \in R \mid \theta)$, with $\theta \in \Psi$, be the power function of a hypothesis test with rejection region $R$ (similarly, $\pi^{'}$ for the rejection region $R^{'}$). The following definition, restated from \cite{KULL:97}, completes the description of the IMP test.

\begin{definition}
\label{def:impt}
For a particular significance level $\alpha$, a test is IMP with respect to a partition $\{A_j\}$ of $W$ and a partition $\{R_j\}$
of the critical region $R$, if for each $A_i$ there are no sets $R^{'}$ and $\{R^{'}_j\}$ such that
\\
1. $R_j=R^{'}_j$ for all $j\neq i$;
\\
2. $P( w \in R^{'} \mid H_0)=\pi^{'}(\theta)=\alpha$, if $\theta \in$ $\overline{W}$;
\\
3. $P( w \in R^{'}_i \mid (Z,\theta_0,\theta_Z))>P( w \in R_i \mid (Z,\theta_0,\theta_Z))$ for any $(Z,\theta_0,\theta_Z) \ \in A_i$.
\end{definition}

Definition \ref{def:impt} implies that no other test of size $\alpha$ exists,  such that the three conditions above can be simultaneously true. 

Now, let $A_Z=\{(Z,\theta_Z,\theta_0): \theta_Z>\theta_0\}$ and $A_0=\{(Z,\theta_Z,\theta_0): \theta_Z=\theta_0\}$. Let $R_Z$, a subset  in a partition of $R$ described by the intersection of the critical region $R$ and a subset of the sample space in which $Z$ is the most likely cluster.
 However, let $R^{'}_Z$ be any subset in a partition of $R^{'}$. We now enunciate Kulldorff's Theorem 2 for the case of the
 Scan-{\mbox ZIP} statistic.

\begin{theorem}
The test based on $\lambda$ forms an IMP test with respect to partitions $\{ A_Z\}$ and  $\{ R_Z\}$.
\end{theorem}
The proof for Theorem 2 can be found in the appendix.

\section{Simulation study}

In order to test our methodology, we constructed different artificial clusters -- which we will call \textit{true} clusters -- using a map consisting of 203 hexagonal cells arranged in a regular grid, each cell with population 1,000. Thus, the total population in the whole map is 203,000. In this framework the distances from any cell to its neighbors are always the same. So, to avoid any tie when measuring distances between any two cells, we introduced a random disturbance in the $x$ and $y$ coordinates of the cells' centroids.

For each cluster, the cases (e.g., a given disease, crime, etc.) are randomly distributed over the map according to a multinomial distribution whose probabilities are proportional to the relative risks assigned to each region. These risks are computed following the procedure described by \cite{KTP:03}, assigning high risks to the regions inside the cluster and lower risks to the remaining regions. The risks of regions inside the cluster are set high enough so that conditioned on the total number of cases, the null hypothesis would be rejected with probability 0.999 when using a binomial test.

A structural zero inside the cluster can be interpreted as a high risk area whose case count is unavailable. In this sense, if there are any regions inside the cluster with structural zeros, then the attributed risk is also high, but the total number of cases is distributed only among the non-structural zero regions of the map. If it happens that a case is assigned to a structural zero region, than that case is rejected and another one is generated. Thus, the case counts of the structural zero regions are always set to zero.

The referred random distribution of cases is repeated thousands of times (say $N$ times), generating thousands of simulated data sets which are used to compute power, sensitivity, and positive predicted value (PPV).

Let us denote the cluster detection algorithms as Scan-Poisson (Kulldorff's standard version of the Poisson-scan algorithm), Scan-\mbox{ZIP} (when the structural zeros are known), and Scan-\mbox{ZIP}+EM (when the structural zeros are unknown). All the three algorithms were used along with circular windows to find the most likely cluster, as described in the following scheme:

\begin{enumerate}
	\item Let $(j)$ be the $j$-\textit{th} nearest region to region $i$, including region $i$ itself, and $z_i^{(j)}$ be the set of the $j$ nearest regions to region $i$. That is, $z_i^{(1)}=\{i\}$, $z_i^{(2)}=\{i,(2)\}$, and so on.
	\item For $j=1, ..., k$, and while the population inside $z_i^{(j)}$ is not bigger than half the total population of the map, compute the likelihood ratio $LR(z_i^{(j)})$.
	\item Steps 1-2 are repeated for $i=1, \ldots, k$.
	\item The most likely cluster is the one corresponding to the maximum of the computed $LR$s.
\end{enumerate}

For each of the $N$ random distributions of cases over the map, test statistic $\lambda_i$, $i=1,\dots, N$, is computed by running the chosen cluster detection algorithm. Then, the obtained $\lambda_i$s are compared to critical value $\lambda^*$, obtained by the Monte Carlo simulation under $H_0$. The power is then given by

$$
\mbox{Power}=\displaystyle \frac{\sum_{i=1}^{N} I(\lambda_i>\lambda^*)}{N},
$$
where $I(\lambda_i>\lambda^*)=1$ if $\lambda_i>\lambda^*$ and 0 otherwise. Sensitivity and PPV are defined in terms of the population of the true and detected clusters \citep{KHK:09} as

\small 
$$
\mbox{Sensitivity} = \frac{\mbox{Pop(Detected Cluster} \cap \mbox{True Cluster)}}{\mbox{Pop(True Cluster)}},
$$

\small 
$$
\mbox{PPV} = \frac{\mbox{Pop(Detected Cluster} \cap \mbox{True Cluster)}}{\mbox{Pop(Detected Cluster)}}.
$$ \normalsize
where Pop() denotes the population. The reported values of sensitivity and PPV are the average values of these proportions along the $N$ Monte Carlo simulations.
\subsection{Simulation design}
\begin{figure}[!ht]
\centering
\includegraphics[width=\linewidth]{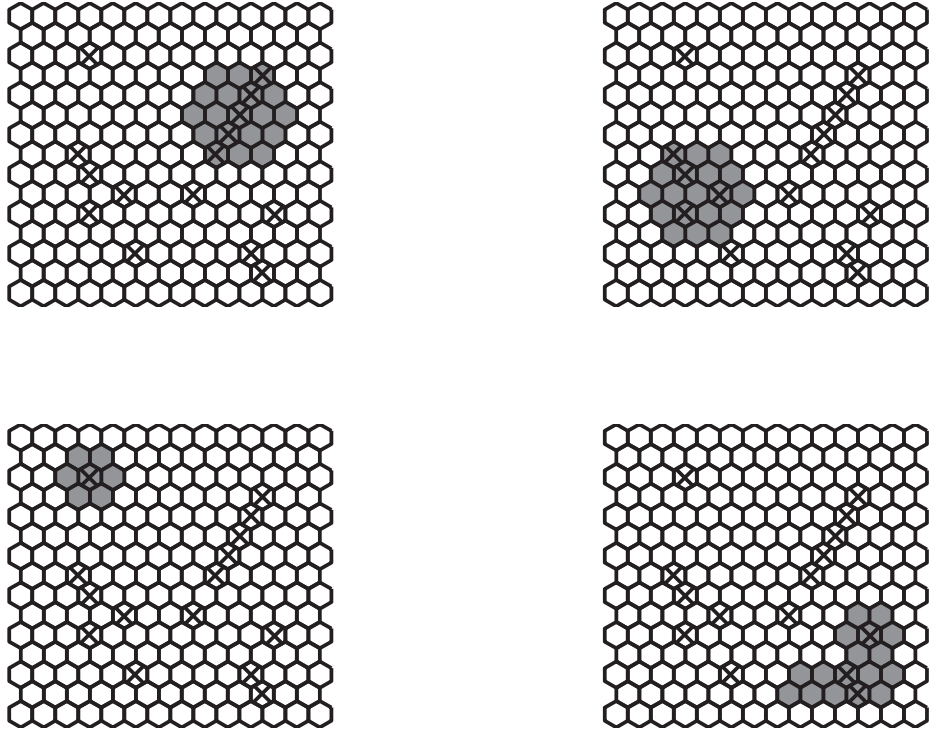}
\caption{Artificial scenarios $A$ (top-left), $B$ (top-right), $C$ (bottom-left), and $D$ (bottom-right). Regions in gray represent the cluster and the $\times$s indicate structural zeros.}
\label{fig:artificial}
\end{figure}

The three algorithms were used to detect the artificial clusters $A$, $B$, $C$, and $D$, which are described below and illustrated in Figure \ref{fig:artificial}. 
\begin{itemize}
	\item Scenario $A$ is a circular cluster crossed by a set of contiguous regions with structural zeros disconnecting the cluster;
	\item Scenario $B$ is a circular cluster with ``random'' structural zeros;
	\item Scenario $C$ is a small circular cluster with a single structural zero region in the middle; and
	\item Scenario $D$ is an irregular L-shaped cluster with a structural zero in the middle of the upper 
	part and with two structural zeros disconnecting the lower left part.
\end{itemize}

Notice, from Figure \ref{fig:artificial}, that in the four scenarios described above, the structural zero regions are the same, and thus the scenarios differ only by the locations of the high-risk regions. These scenarios were designed in order to simulate realistic situations: a cluster disconnected by a systematic pattern ($A$), random lack of information ($B$), a cluster with a hole ($C$), and a problematic irregular cluster ($D$).

\begin{figure}[!ht]
\centering
\includegraphics[width=\linewidth]{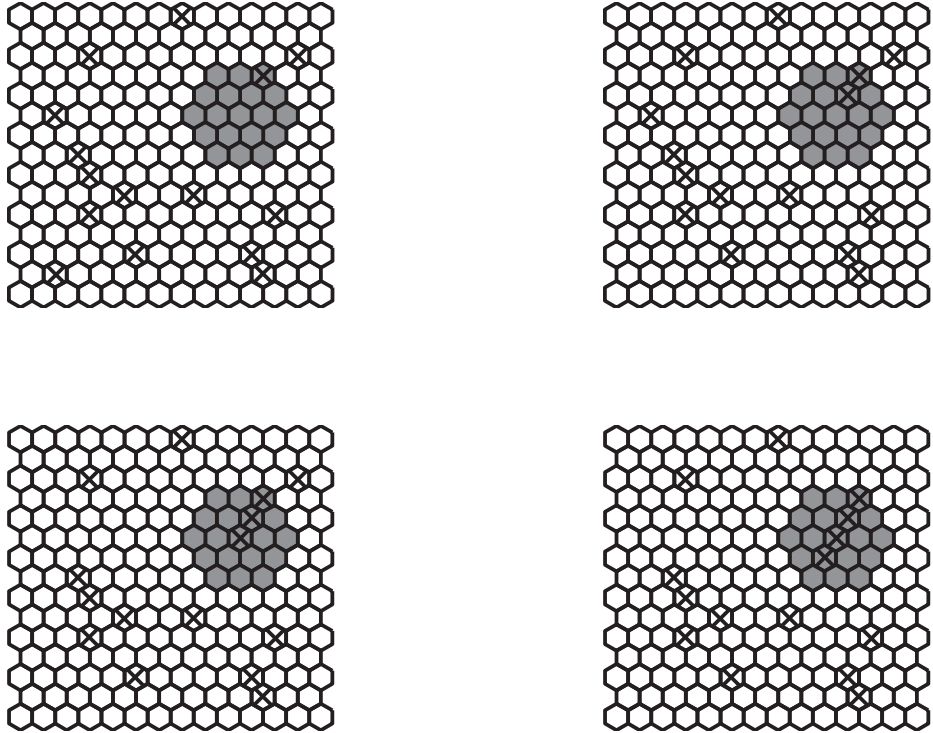}
\caption{Artificial progressive scenarios $A_1$ (top-left), $A_2$ (top-right), $A_3$ (bottom-left), and $A_4$ (bottom-right). Regions in gray represent the cluster and the $\times$s indicate structural zeros.}
\label{fig:prog}
\end{figure}

In order to illustrate the deteriorating performance of the Scan-Poisson algorithm as we progressively add structural zeros to the true cluster, we designed four extra {\it progressive scenarios}, $A_1$--$A_4$ (see Figure \ref{fig:prog}). These  scenarios are modified versions of scenario $A$, in the sense that high-risk regions are the same, and thus that these scenarios differ only by the locations of the structural zero regions. Starting from scenario $A_1$, structural zeros are gradually moved from outside to inside the true cluster, thereby preserving the total number of structural zeros in the map. Further, we have included scenario $A_0$, also designed from scenario $A$, but with no structural zeros.

All scenarios have 15 structural zero regions (except for $A_0$) with the total number of cases $M$ fixed at 0.25\% of the total population: 507 cases. For each scenario we run 10,000 Monte Carlo replications under the null hypothesis for each method to compute the critical value, 
and another 10,000 replications under the alternative hypothesis to compute power, sensitivity, and PPV. The random distributions of cases are always the same for the three methods.

\subsection{Analysis for the simulated data}
The results obtained using the simulated data are presented in Table \ref{tab:results}, in terms of power, sensitivity, and PPV, for the three different methods applied to the nine designed scenarios. As we can observe, regarding the estimates of power, sensitivity, and PPV, both Scan-\mbox{ZIP} and Scan-\mbox{ZIP}+EM algorithms are systematically superior to the Scan-Poisson algorithm, the last one being comparable only when the true cluster presents none or few structural zeros. Besides, the performance of Scan-\mbox{ZIP}+EM is very close to that of Scan-\mbox{ZIP}.

\begin{table}[!hb]
\begin{center}
\caption{Power, sensitivity and positive predictive value obtained by the three methods for artificial clusters  $A_0$, $A$, $B$, $C$ and $D$ and progressive $A_1$--$A_4$.}
\label{tab:results}
\scriptsize
\begin{tabular}{@{\extracolsep{-1.2pt}}cccc|ccc|ccc}\hline
\multirow{2}{*}{\textbf{Scen.}} 
& \multicolumn{3}{c}{\textbf{Scan-Poisson}} & \multicolumn{3}{c}{\textbf{Scan-\mbox{ZIP}}} 
& \multicolumn{3}{c}{\textbf{Scan-\mbox{ZIP}+EM}}\\ \cline{2-10}
 & \textbf{Power} 
& \textbf{Sens.} & \textbf{PPV} &  \textbf{Power} & \textbf{Sens.} & \textbf{PPV} 
& \textbf{Power} & \textbf{Sens.} & \textbf{PPV}\\ \hline
$A_0$ & 0.9502 & 0.8533 & 0.8737 & 0.9502 & 0.8533 & 0.8737 & 0.9554 & 0.8599 & 0.8466\\\hline 
$A$ & 0.6758 & 0.6496 & 0.7475 & 0.9479 & 0.8664 & 0.8718 & 0.9152 & 0.8400 & 0.8417\\
$B$ & 0.6361 & 0.4828 & 0.6915 & 0.9500 & 0.8671 & 0.8796 & 0.9149 & 0.8426 & 0.8578\\
$C$ & 0.8672 & 0.8999 & 0.7593 & 0.9521 & 0.9339 & 0.8935 & 0.9369 & 0.9265 & 0.8660\\
$D$ & 0.6769 & 0.5337 & 0.6304 & 0.8756 & 0.6938 & 0.6534 & 0.8376 & 0.6880 & 0.6421\\\hline
$A_1$ & 0.9533 & 0.8070 & 0.8154 & 0.9533 & 0.8533 & 0.8575 & 0.9453 & 0.8483 & 0.8232\\
$A_2$ & 0.9031 & 0.7359 & 0.8011 & 0.9475 & 0.8587 & 0.8642 & 0.9311 & 0.8455 & 0.8313\\
$A_3$ & 0.8176 & 0.7107 & 0.7545 & 0.9518 & 0.8690 & 0.8702 & 0.9297 & 0.8532 & 0.8451\\
$A_4$ & 0.6909 & 0.6228 & 0.7033 & 0.9508 & 0.8785 & 0.8762 & 0.9152 & 0.8545 & 0.8537\\\hline
\end{tabular}
\normalsize
\end{center}
\end{table}

Regarding the progressive scenarios, the Scan-Poisson algorithm steadily deteriorates as the number of structural zeros increases. The Scan-\mbox{ZIP} algorithm is able to maintain its good properties irrespective of the number of structural zeros, while the Scan-\mbox{ZIP}+EM algorithm deteriorates much slower than the Scan-Poisson algorithm, remaining close to the Scan-ZIP algorithm. This suggests that the structural zero estimates obtained by the EM procedure are quite reasonable and that the Scan-\mbox{ZIP}+EM algorithm is clearly capable of capturing the presence of structural zeros.

The results obtained for scenario $A_0$ show that the lack of structural zeros does not compromise the performance of the Scan-\mbox{ZIP}+EM method, as its performance is very close to that of the Scan-Poisson algorithm. This indicates that the proposed statistic can be possibly used as a more general model than the standard Poisson approach. For this scenario, the results for the Scan-\mbox{ZIP} method are obviously identical to those of the Scan-Poisson algorithm.

\subsection{Discussing the Scan-Poisson type I error}
It is important to study the performance of the discussed methods for maps that include structural zero regions but with nonexistent clusters. To that purpose, we conducted another experiment in which we considered 15 structural zero regions, and according to the ZIP model, we randomly distributed 507 cases over the map under the null hypothesis. This experiment was repeated 10,000 times. In this setting, the expected number of cases was the same for all regions, except for the regions presenting a structural zero. We then applied the three scan methods to each one of the 10,000 generated samples in order to compute the probability of type I error for each approach. 

For $LLR$ tests with $\alpha=0.05$, the Scan-ZIP and Scan-ZIP+EM methods presented type I error rates of 0.0548 and 0.0561, respectively. Thus, both methods produced error rates close to the nominal significance level. However, the Scan-Poisson type I error proportion was 0.1006. 

The Scan-Poisson performance was poorer  because the simulated process under $H_0$  was not a homogeneous one, an assumption required for building the Scan-Poisson $LLR$ empirical distribution. This violation in the homogeneity assumption caused a substantial bias in the $LLR$ estimated distribution. As a consequence, the Scan-Poisson method was more likely to detect spurious significant clusters than it would be expected in normal conditions. This problem obviously did not affect the Scan-ZIP method that proved to be more robust than the Scan-Poisson method to the occurrences of type I error. The Scan-ZIP+EM method is apparently able to adequately estimate the model. 

\section{Application}
In many practical situations, we do not actually know whether or not the data include structural zeros. However, the \mbox{ZIP} model can be used to account for excess zeros when the practitioner suspects that there are more zeros than that expected for count data under the assumption of a Poisson distribution. For example, in small towns, due to the lack of a health center designed to diagnose a given disease, the number of reported cases may be set as zero. Further, limited health insurance over community members may prevent access to health center facilities, thereby causing a downward bias in the notification statistics provided by this health center. 

Another source of negative bias in the notification statistics is due to under-reporting. This may occur despite state and local laws requiring medical providers to report notifiable cases to public health authorities and can be attributed to incomplete reporting by the health staff. We would like to stress the fact that in equation (\ref{eq1}), the probability of zero cases includes two possible sources: the zero that occurs due to a random or stochastic situation and the zero that is due to the structural (and almost deterministic) nature of the event.

\subsection{Breast cancer deaths in northern Brazil}

\begin{figure}[!ht]
\centering
\includegraphics[width=\linewidth]{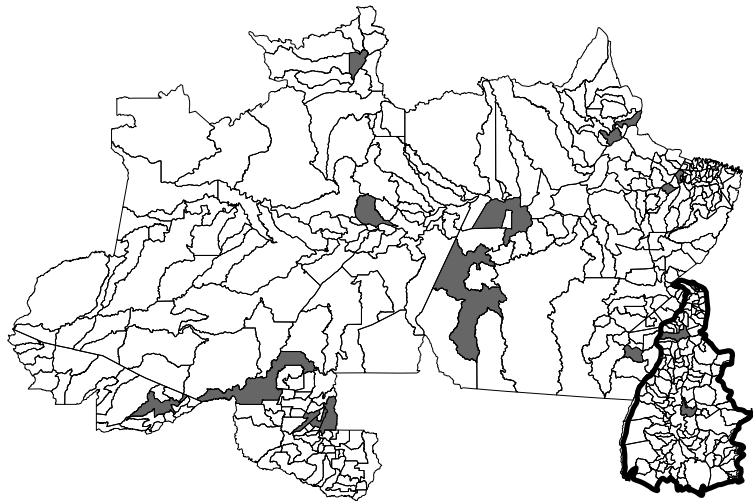}
\caption{Northern region of Brazil. Gray regions indicate municipalities with at least one death due to breast cancer.}
\label{fig:amazon}
\end{figure}

In order to illustrate some of these points, consider the northern area of Brazil, divided into municipalities. For each municipality the number of breast cancer deaths from 2008 to 2011 are recorded. Areas with non-zero counts are then displayed in a map. We observe that most of these areas in the map (see Figure \ref{fig:amazon}) present zero cases. This has nothing to do with possible immunity to this disease. Actually, that is just an artifact caused by the absence of mammographers in these areas.

For example, according to the Brazilian National Cancer Institute (INCA), in 2011, only 86 mammographers were available for the entire northern region that includes a total of 450 municipalities. Furthermore, the northern region of Brazil is an enormous territory (approximately 40$\%$ of the US area), with a very low population density and encompassing most of the Amazon forest, thus making mobility very difficult. Further, given an imperfect system of notification of cases (e.g., not pointing out the municipality of residence), and we will certainly observe ``false'' zeros, caused by a series of external factors. As we can observe from Table \ref{tab:municipalities}, only 16 of the 450 municipalities reported some notification records 
concerning deaths caused by breast cancer. 

\begin{table}[!hb]
\caption{Number of breast cancer death notifications in the northern area of Brazil from 2008 to 2011.}
\label{tab:municipalities}
\begin{center}
\begin{tabular}{lrrrrr}\hline
& \multicolumn{4}{c}{\textbf{Year}} & \\	
Municipality        &	2008	& 2009	& 2010	& 2011	& Total	\\	\hline
Porto Velho         & 1	 & -	& 7	 & 13	& 21 \\
Rio Branco          & 1	 & 6	& 2	 & 5	& 14 \\
Manaus	            & 11 & 12	& 12 & 22	& 57 \\
Boa Vista	          & -	 & 2	& 9	 & 3	& 14 \\
Belém	              & 21 &54	& 55 & 51	& 181 \\
Santarém	          & -	 & 2	& 2	 & 5	& 9 \\
Macapá¡	            & -	 & 6	& 8	 & 8	& 22 \\
Araguaína	          & 15 & 9	& 12 & 6	& 42 \\
Palmas	            & 1	 & 5	& 5	 & 5	& 16 \\ 
Augustinópolis	    &  - & 	-	& -	 & 1	& 1\\
Redenção	          & -	 & -	& 1	 & -	& 1\\
Itaituba	          & 1	 & -	& -	 & -	& 1\\
Abaetetuba	        & -	 &-	  & -	 & 1	& 1\\
Ouro Preto do Oeste &	-	 & -	& 1	 & -	& 1\\
Jaru	              & -	 & 1	& -	 & -	& 1\\
Ji-Paraná¡	          & 1	 & 1	& -	 & -	& 2\\
Other municipalities& -  & -  & -  & -  & - \\ \hline
Total               &	52 & 98	& 114& 120& 384\\	\hline
\end{tabular}
\end{center}
\end{table}

We applied our methods to the Tocantins state (highlighted in Figure \ref{fig:amazon}). According to the recorded data, Tocantins presented 14 deaths by breast cancer during 2008, for a population at risk of 690,104 women. We considered that municipalities with no available mammographers present structural zeros. In fact, we observed that all municipalities without functional mammographers presented no records of deaths by breast cancer during 2008. 

\begin{figure}[!ht]
\centering
\includegraphics[width=\linewidth]{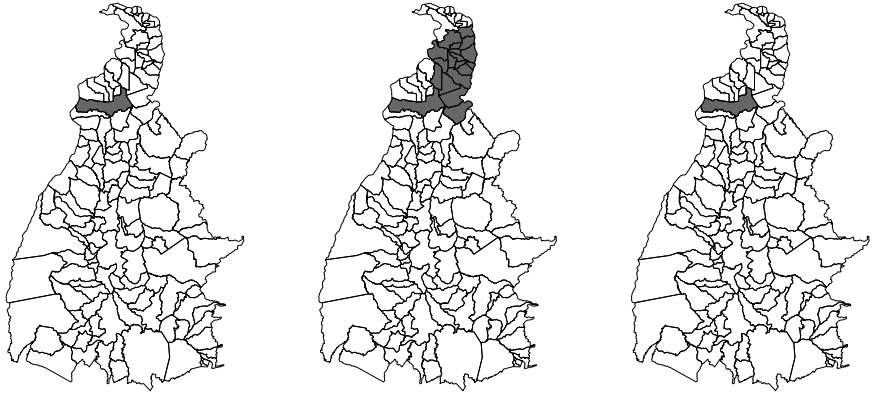}
\caption{Most likely clusters of breast cancer deaths found by the Scan-Poisson (left), Scan-ZIP (center), and Scan-ZIP+EM (right) method in the state of Tocantins, Brazil.}
\label{fig:tocantins}
\end{figure}

The most likely clusters detected by the Scan-Poisson, Scan-ZIP, and Scan-ZIP+EM methods are shown in Figure \ref{fig:tocantins}. The cluster detected by the Scan-Poisson method, including only one municipality (Araguaína), is significant ($p$-value = 0.0021), while the most likely clusters detected by the other methods are not: $p$-value = 0.7712 for the Scan-ZIP and 0.1750 for the Scan-ZIP+EM method. The non-significant cluster detected by the Scan-ZIP method includes the municipality of AraguaÃ­na and some of its neighbours. 
\\ \\
The disagreement between the Scan-Poisson method and the other methods can be justified by the results of our last simulations: as we mentioned at the end of Section 5, the Scan-Poisson method presents some bias towards the detection of spurious significant clusters when structural zeros are present in the data. These findings reinforce the importance of our proposed methodology as a useful tool in preventing incorrect identification of clusters when zero-inflation is present in the data.

\subsection{Oral cancer cases in Georgia, United States}

\begin{figure}[!ht]
\centering
\includegraphics[width=\linewidth]{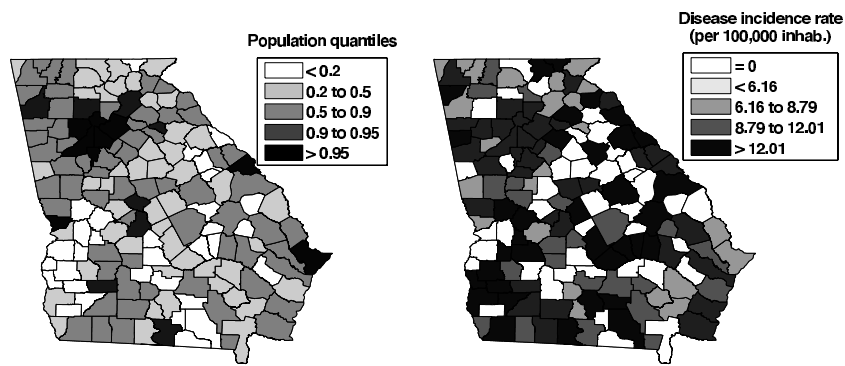}
\caption{Population distribution (left) and oral cancer death rates (right) in Georgia, United States, between 1994 and 1995.}
\label{fig:pop-rate}
\end{figure}

We now present an application to oral cancer cases data in Georgia, United States. According to the International Classification of Diseases, oral cancer is a malignant neoplasm of lip, oral cavity, or pharynx, and can be described as an abnormal, malignant tissue growth in the mouth. The data set used in our application was obtained at {\tt  http://oasis.state.ga.us} and consists of 430 deaths by oral cancer between 1994 and 1995, for all races, ethnicities, and life stages. The resulting map is divided into 159 counties and the population at risk is 7,157,165. 
Figure \ref{fig:pop-rate} shows the population distribution over the map of Georgia and the disease incidence rates. Our guess is that the data may be affected by under-reporting and possible presence of structural zeros. This can be, for example, due to the lack of health insurance coverage among some citizens in Georgia, thereby preventing them from obtaining any kind of help, which leads to there being no diagnosed cause of death. Another possibility is under-reporting, for example, due to failures in filling medical protocols by the medical staff. 

\begin{figure}[!ht]
\centering
\includegraphics[width=\linewidth]{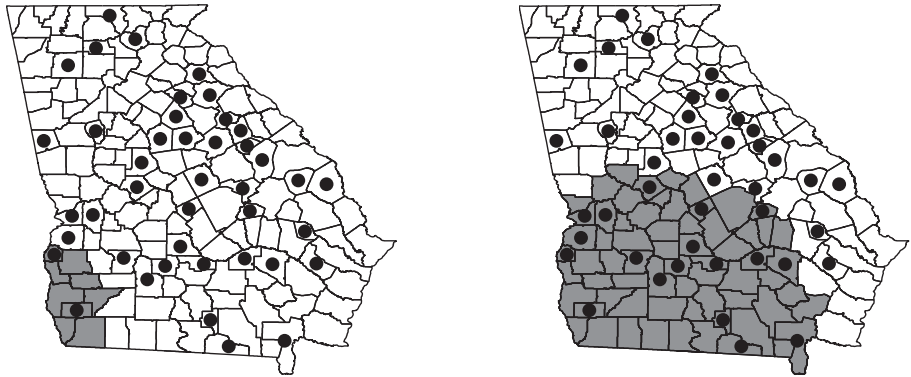}
\caption{Clusters found by the Scan-Poisson (left) and Scan-ZIP+EM (right) algorithms. Black dots indicate zero case counts.}
\label{fig:clusters}
\end{figure}

We then run the cluster detection algorithms using the Poisson and ZIP models. Since we have no information about the location of possible structural zeros, we used the Scan-\mbox{ZIP}+EM version for the ZIP case. The most likely clusters found by each method are shown in Figure \ref{fig:clusters}. The black dots indicate regions with zero case counts. The cluster produced by the Scan-Poisson algorithm is not significant at $\alpha = 0.05$ ($p\mbox{-value} =0.056$) while the cluster detected by the Scan-\mbox{ZIP}+EM algorithm is highly significant ($p\mbox{-value}=0.002$).  Besides, the cluster produced by the Scan-Poisson algorithm differs substantially from the one that was produced by the Scan-\mbox{ZIP}+EM algorithm, and the former seems to misrepresent a high-incidence oral cancer area.

We notice from Figure \ref{fig:clusters} that the presence of some regions with zero case counts prevents the Scan-Poisson method from finding a solution (or cluster) that includes the high-incidence areas in the lower-middle part of the map. This happens because the inclusion of these regions lowers the test statistic ($\lambda$). However, the solution provided by the Scan-ZIP+EM algorithm is very robust  in the sense that it does include several zero case-count areas, since their impact in reducing $\lambda$ is easily compensated by the inclusion of the high incidence-regions.

\section{Discussion}
In this article, we focus on extensions of the scan statistic proposed by \cite{KULL:97}  for cluster detection of spatial count data. The new scan statistics, denoted as Scan-\mbox{ZIP} and Scan-\mbox{ZIP}+EM, allow for the occurrence of zero-inflated spatial data. Our simulations revealed that the Scan-Poisson statistic proposed by \cite{KULL:97} steadily deteriorates as the number of structural zeros increases, producing biased inferences. This can be evaluated in terms of power, sensitivity, and PPV measures. Our simulations also revealed that the Scan-\mbox{ZIP} and Scan-\mbox{ZIP}+EM statistics are, most of the time, either superior or comparable to the Scan-Poisson statistic for all scenarios. When the structural zeros are known, Scan-\mbox{ZIP} is far superior to the other algorithms. The real data application reinforces these conclusions: for high-incidence areas, which nonetheless present zero case counts, the Scan-Poisson method fails to provide a reasonable solution or cluster. 

Another important finding shown in our simulations is that when the data include structural zeros, the Scan-Poisson method is more likely to detect spurious significant clusters than the Scan-\mbox{ZIP} and Scan-\mbox{ZIP}+EM methods.

Like Kulldorff's scan statistic, the ZIP scan statistic can be linked to more sophisticated cluster detection techniques, allowing for the detection of irregularly shaped clusters. Moreover, it seems quite straightforward to make extensions to space-time and prospective space-time versions. However, it should be borne in mind that the computational effort required by the EM algorithm increases as the number of candidate clusters increases, since we must estimate structural zeros for each candidate cluster. In our experiments, the Scan-ZIP+EM was about eight times slower than the Scan-ZIP and Scan-Poisson.

In this work, we used area-aggregated data for simulation and applications. With the formulation presented in this paper, the proposed method is restricted to this type of data. For instance, our methodology is not suitable for point pattern data. In the future, a more general approach could be developed to account for point pattern data sets.

In Section 3, we assume that the case counts in the regions follow independent ZIP random variables with the same probability $p$ of a structural zero. However, following \cite{Lambert:92},  a more flexible model can be devised by considering the case where probability $p_i$ of a structural zero in region $i$ is modelled as a function of covariates using a logistic regression.

\section*{Acknowledgements}
André L. F. Cançado was partially supported by DPP/UnB. Cibele Q. da-Silva was supported by the National Research Council (CNPq-Brazil, BPPesq) and by the Office to Improve University Research (CAPES-Brazil) via Project PROCAD-NF 2008.

\footnotesize

\bibliographystyle{plainnat}      
\bibliography{zip}   

\normalsize

\section*{Appendix}
\begin{proof}{\textit{of Theorem 1}} Let $\lambda(D)$ and $\lambda(D^{'})$ denote the values for the test statistic
for the two different data sets. Further, consider
\begin{eqnarray*}
I(Z)=\left[\frac{\sum_{i \in {Z}}^{}x_i(1-d_i)}{\sum_{i \in {Z}}^{}n_i(1-d_i)}\right] \ \ \ \mbox{and} \ \ \
O(Z)=\left[\frac{\sum_{i \notin {Z}}^{}x_i(1-d_i)}{\sum_{i \notin {Z}}^{}n_i(1-d_i)}\right].
\end{eqnarray*}
Similarly, let
\begin{eqnarray*}
I'(Z)=\left[\frac{\sum_{i \in {Z}}^{}x_i'(1-d_i')}{\sum_{i \in {Z}}^{}n_i'(1-d_i')}\right] \ \ \ \mbox{and} \ \ \
O'(Z)=\left[\frac{\sum_{i \notin {Z}}^{}x_i'(1-d_i')}{\sum_{i \notin {Z}}^{}n_i'(1-d_i')}\right],
\end{eqnarray*}
and let

$$C=\sum_{i=1}^{k}x_i(1-d_i) = \sum_{j=1}^{k}x_j'(1-d_j'),$$

$$R=\sum_{i=1}^{k}n_i(1-d_i)=\sum_{j=1}^{k}n_j'(1-d_j'),$$

$$c(Z)=\sum_{i \in {Z}}^{}x_i(1-d_i),$$

$$r(Z)=\sum_{i \in {Z}}^{}n_i(1-d_i),$$

$$c'(Z)=\sum_{i \in {Z}}^{}x_i'(1-d_i'),$$

$$r'(Z)=\sum_{i \in {Z}}^{}n_i'(1-d_i'),$$

$$K=\left[\frac{\sum_{i=1}^{k}x_i(1-d_i)}{\sum_{i=1}^{k}n_i(1-d_i)}\right]^{\sum_{i=1}^{k}x_i(1-d_i)}.$$

As stated previously, these quantities are related to the sufficient statistics of the MLEs.
\\ \\ 
Under the null hypothesis, $\lambda(D)=1,$ and this implies that 
\begin{eqnarray*}
K&=&\mbox{sup}_{Z \in \mathcal{Z}} \left[I(Z)\right]^{c(Z)} \left[O(Z)\right]^{C-c(Z)}
= \left[I(\hat{Z})\right]^{c(\hat{Z})} \left[O(\hat{Z})\right]^{C-c(\hat{Z})} \\
&=& \mbox{sup}_{Z \in \mathcal{Z}} \left[I'(Z)\right]^{c'(Z)} \left[O'(Z)\right]^{C-c'(Z)}
=   \left[I'(\tilde{Z}^{'})\right]^{c'(\tilde{Z}^{'})} \left[O'(\tilde{Z}^{'})\right]^{C-c'(\tilde{Z}^{'})}.
\end{eqnarray*}
Thus, $c(\hat{Z})=C=c'(\tilde{Z}^{'})$, and the distributions of $\lambda(D)$ and $\lambda(D^{'})$ are the same.
\\ \\
Under the alternative hypothesis $\lambda(D)>1$, and we need to show that $\lambda(D') \geq \lambda(D)$.
Notice that under the conditions of the theorem, $c'(\tilde{Z}^{'}) \geq c(\hat{Z})$, since 
$$\sum_{i \in \tilde{Z}^{'}}^{}x_i'(1-d_i')= \sum_{i \in \hat{Z}}^{}x_i'(1-d_i')+
\sum_{i \in \overline{\hat{Z}}\cap \tilde{Z}^{'}}^{}x_i'(1-d_i')
\geq \sum_{j \in \hat{Z}} x_j(1-d_j),$$
as $\overline{\hat{Z}}\cap \tilde{Z}^{'}$ might be different from the empty set. 
When $\lambda(D) > 1$, we have from equation (\ref{eq9}) that
\begin{eqnarray*}
\nonumber \lambda(D) 
&=& \mbox{sup}_{Z \in \mathcal{Z}} \frac{1}{K} \left[I(Z)\right]^{c(Z)} \left[O(Z)\right]^{C-c(Z)}\\
&=& \frac{1}{K} \left[I(\hat{Z})\right]^{c(\hat{Z})} \left[O(\hat{Z})\right]^{C-c(\hat{Z})} \\
&=& \frac{1}{K} \left[\frac{c(\hat{Z})}{r(\hat{Z})}\right]^{c(\hat{Z})} \left[\frac{C-c(\hat{Z})}{R-r(\hat{Z})}\right]^{C-c(\hat{Z})}\\
&\leq& \frac{1}{K}\left[\frac{c'(\hat{Z})}{r'(\hat{Z})}\right]^{c'(\hat{Z})}
\left[\frac{C-c'(\hat{Z})}{R-r'(\hat{Z})}\right]^{C-c'(\hat{Z})}\\
&\leq & \mbox{sup}_{Z \in \mathcal{Z}} \frac{1}{K} \left[I'(Z)\right]^{c'(Z)} \left[O'(Z)\right]^{C-c'(Z)}\\
&=& \frac{1}{K} \left[I'(\tilde{Z}^{'})\right]^{c'(\tilde{Z}^{'})} \left[O'(\tilde{Z}^{'})\right]^{C-c'(\tilde{Z}^{'})}=  \lambda(D^{'}).
\end{eqnarray*}
The first inequality holds since for any constants $\alpha$, $\beta$, and $N$,
$(\alpha n)^n(\beta(N-n))^{N-n}$ is an increasing function of $n$ when $\alpha n > \beta(N-n)$. 
This is true since $\lambda(D) > 1$ implies that $I(\hat{Z})>O(\hat{Z})$, that is, $\frac{c(\hat{Z})}{r(\hat{Z})}>\frac{C}{R}$. 
This also means that $I'(\hat{Z}) > O'(\hat{Z})$. In order to verify this, using a proof by contradiction, let us suppose 
that $I'(\hat{Z}) \leq O'(\hat{Z})$. Then, $\frac{c'(\hat{Z})}{r'(\hat{Z})}\leq\frac{C}{R}$. 
Since $c'(\hat{Z}) \geq c(\hat{Z})$, this implies that 
$\frac{c(\hat{Z})}{r(\hat{Z})} \leq \frac{c'(\hat{Z})}{r(\hat{Z})} \leq \frac{C}{R}$ \ whenever \ $r(\hat{Z})=r'(\hat{Z})$, 
which is absurd.
\end{proof}

\begin{proof}{\textit{of Theorem 2}} According to Definition \ref{def:impt}, in order to prove that $\lambda$ is an
IMP test, it is necessary to show that if statements (1) and (2)
are true, then (3) cannot hold. This is equivalent to showing that for any $(Z,\theta_0,\theta_Z) \ \in A_Z$,
\begin{eqnarray}
\label{ineq1}
P( w \in R^{'}_Z \mid (Z,\theta_0,\theta_Z))-P( w \in R_Z \mid (Z,\theta_0,\theta_Z))\leq 0.
\end{eqnarray}
For an arbitrary $Z$, let $D_{-}=\{w: w \in R_Z, w \notin R^{'}_Z \}$ and 
$D_{+}=\{w: w \in R^{'}_Z, w \notin R_Z \}$. Define
$$ M = \mbox{sup}_{ \ w \in D_{+}} \frac{L(Z,\theta_Z,\theta_0 \mid w)}{L(\theta_0 \mid w)}.$$
By the definition of $D_{+}$ and $D_{-}$, since $R_Z$ is described in terms of $Z$,
which is the most likely cluster in a subset of the sample space, we have that
each $w$ in $D_{-}$ has a higher likelihood ratio than any $w$ in $D_{+}$; that is, 
$$ M = \mbox{sup}_{ \ w \in D_{+}} \frac{L(Z,\theta_Z,\theta_0 \mid w)}{L(\theta_0 \mid w)} \leq
\mbox{inf}_{ \ w \in D_{-}} \frac{L(Z,\theta_Z,\theta_0 \mid w)}{L(\theta_0 \mid w)},$$
\begin{eqnarray*}
M &=&\mbox{sup}_{ \ w \in D_{+}}
\frac{\left[\frac{\sum_{i \in Z_{w}}^{}x_i(1-d_i)}{\sum_{i \in Z_{w}}^{}n_i(1-d_i)}
\right]^{\sum_{i \in Z_{w}}^{}x_i(1-d_i)}
\left[\frac{\sum_{j \notin Z_{w}}^{}x_j(1-d_j)}{\sum_{j \notin Z_{w}}^{}n_j(1-d_j)}
\right]^{\sum_{j \notin Z_{w}}^{}x_j(1-d_j)}}
{\left[\frac{\sum_{i=1}^{k}x_i(1-d_i)}{\sum_{i=1}^{k}n_i(1-d_i)}
\right]^{\sum_{i=1}^{k}x_i(1-d_i)}}\\ \\
&\leq & \mbox{inf}_{ \ w \in D_{-}}
\frac{\left[\frac{\sum_{i \in Z_{w}}^{}x_i(1-d_i)}{\sum_{i \in Z_{w}}^{}n_i(1-d_i)}
\right]^{\sum_{i \in Z_{w}}^{}x_i(1-d_i)}
\left[\frac{\sum_{j \notin Z_{w}}^{}x_j(1-d_j)}{\sum_{j \notin Z_{w}}^{}n_j(1-d_j)}
\right]^{\sum_{j \notin Z_{w}}^{}x_j(1-d_j)}}
{\left[\frac{\sum_{i=1}^{k}x_i(1-d_i)}{\sum_{i=1}^{k}n_i(1-d_i)}
\right]^{\sum_{i=1}^{k}x_i(1-d_i)}}\\ \\
&=&\mbox{inf}_{ \ w \in D_{-}} \frac{L(Z,\theta_Z,\theta_0 \mid w)}{L(\theta_0 \mid w)}.
\end{eqnarray*}
The proof of inequality (\ref{ineq1}) for any $(Z, \theta_Z, \theta_0) \in A_Z$ follows largely from \cite{KULL:97}, where it is verified that 
\scriptsize
\begin{eqnarray*}
P( w \in R^{'}_Z \mid (Z,\theta_0,\theta_Z))-P( w \in R_Z \mid (Z,\theta_0,\theta_Z))
\leq M(P(w \in R^{'} \mid H_0) - P(w \in R \mid H_0))=0.
\end{eqnarray*}
\normalsize
The last equality holds since $R_j=R_j^{'}$ for all $j \neq Z$, 
according to statement 2 in Definition \ref{def:impt}.
\end{proof}

\end{document}